\documentclass[12pt]{article}

\usepackage{amsmath, amssymb, mathtools}
\usepackage{type1cm}
\usepackage{mathrsfs}
\usepackage{physics}
\usepackage[T1]{fontenc}
\usepackage{fancyhdr}
\usepackage{appendix}
\usepackage{titlesec}
\usepackage{cite}
\usepackage{color}
\usepackage[top=25truemm,bottom=25truemm,left=20truemm,right=20truemm]{geometry}
\usepackage[dvipdfmx]{graphicx}

\baselineskip=\normalbaselineskip

\newcommand{\beq}{\begin{equation}}
\newcommand{\eeq}{\end{equation}}
\newcommand{\beqn}{\begin{equation*}}
\newcommand{\eeqn}{\end{equation*}}

\newcommand\als[1]{\begin{align}\begin{split}#1\end{split}\end{align}}

\makeatletter
 
 \@addtoreset{equation}{section}
\makeatother

\begin{document}
\setcounter{footnote}{0}
\setcounter{tocdepth}{3}
\bigskip
\def\thefootnote{\arabic{footnote}}

\begin{titlepage}
\renewcommand{\thefootnote}{\fnsymbol{footnote}}
\begin{normalsize}
\begin{flushright}
\begin{tabular}{l}
UTHEP-754\\
DIAS-STP-20-22
\end{tabular}
\end{flushright}
  \end{normalsize}

~~\\

\vspace*{0cm}
    \begin{Large}
       \begin{center}
         {Laplacians on Fuzzy Riemann Surfaces}
       \end{center}
    \end{Large}

\vspace{0.7cm}

\begin{center}
Hiroyuki A\textsc{dachi}$^{1)}$\footnote[1]
            {
e-mail address : 
adachi@het.ph.tsukuba.ac.jp},
Goro I\textsc{shiki}$^{1),2)}$\footnote[2]
            {
e-mail address : 
ishiki@het.ph.tsukuba.ac.jp},
Satoshi K\textsc{anno}$^{1)}$\footnote[3]
            {
e-mail address : 
kanno@het.ph.tsukuba.ac.jp}
and
Takaki M\textsc{atsumoto}$^{3)}$\footnote[4]
            {
e-mail address : 
takaki@stp.dias.ie}

\vspace{0.7cm}

     $^{ 1)}$ {\it Graduate School of Science and Technology, University of Tsukuba, }\\
               {\it Tsukuba, Ibaraki 305-8571, Japan}\\

     $^{ 2)}$ {\it Tomonaga Center for the History of the Universe, University of Tsukuba, }\\
               {\it Tsukuba, Ibaraki 305-8571, Japan}\\
                                  
     $^{ 3)}$ {\it School of Theoretical Physics, Dublin Institute for Advanced Studies }\\
               {\it 10 Burlington Road, Dublin 4, Ireland}\\
               \end{center}

\vspace{0.5cm}

\begin{abstract}
\noindent
We consider the matrix regularization of scalar fields on a 
Riemann surface with a general gauge-field background. We propose a construction of 
the fuzzy version of the Laplacian.
\end{abstract}

\end{titlepage}

\tableofcontents

\section{Introduction}
The concept of noncommutative geometry naturally arises in superstring theory
\cite{Seiberg:1999vs} 
and is expected to give a wider framework of geometry 
 admitting also theories of quantum gravity.
The matrix models \cite{Banks:1996vh,Ishibashi:1996xs}, which are conjectured to be nonperturbative formulations of 
M-theory and superstring theories, also involve noncommutative geometry 
and various objects such as membranes or D-branes are described 
in terms of fuzzy (finite noncommutative) geometry in the matrix models.

The main purpose of this paper lies in understanding the fuzzy geometry by 
investigating the so-called matrix regularization \cite{deWit:1988wri}.
In particular,  for an arbitrary fuzzy Riemann surface 
with (or without) a general gauge-field background, we give a construction of the 
fuzzy version of the Laplacian, which has rich information on 
the geometry and is needed to study scalar field theories on the fuzzy surface.

The matrix regularization is a method of constructing a fuzzy space from 
a given ordinary commutative space.
This method is very useful, because it enables us to understand elusive 
fuzzy geometry in terms of well-established differential geometry of 
commutative spaces.
For a given compact Riemann surface $M$ with 
a symplectic form $\omega$,
the matrix regularization is defined as a linear map 
$T_N:C^{\infty}(M) \to M_N(\mathbb{C})$ which satisfies \cite{Arnlind:2010ac}
\begin{align}
&\lim_{N\to \infty} |T_N(f)T_N(g)-T_N(fg) | = 0,
\label{MatReg1} \\
&\lim_{N\to \infty} |i\hbar_N^{-1}[T_N(f), T_N(g)]- T_N(\{f,g \})| = 0,
\label{MatReg2} \\
&\lim_{N\to \infty} \hbar_N \mathrm{Tr} \, T_N(f) - \frac{1}{2\pi}\int_M \omega f =0,
\label{MatReg3} 
\end{align}
for any $f,g\in C^{\infty}(M)$.
Here, $\hbar_N=V/N$,  $V=\frac{1}{2\pi}\int_M \omega$,
$\{ \; , \; \}$ is the Poisson bracket defined by $\omega$ 
and $|\cdot|$ is a matrix norm.
The equation (\ref{MatReg1}) states that the algebraic structure 
of functions are well approximated by using the noncommutative matrix algebra 
and the approximation becomes more precise as the matrix size 
$N$ goes to infinity. 
The equation (\ref{MatReg2}) shows that  
the Poisson bracket is approximated by the matrix commutator, 
and thus the matrix regularization can be seen as a generalization of 
the canonical quantization of classical mechanics such that
the phase space is not just a plane but the general compact surface $M$.
The equation (\ref{MatReg3}) is needed to avoid the trivial case, 
$T_N(f)=0$ for any $f$, 
and is essential to derive the actions of the matrix models from 
the worldvolume theories of a membrane or a string \cite{deWit:1988wri}.

The matrix regularization can be explicitly 
constructed by the Berezin-Toeplitz quantization
\cite{Klimek:1992a, Klimek:1992b,Bordemann:1993zv, Ma-Marinescu}. 
In this quantization,
as we will describe in more detail in the next section, one starts from a 
suitably constructed Dirac operator $D$ with totally $N$ normalizable 
zero modes.
Then, one obtains the map $T_N$ satisfying (\ref{MatReg1})--(\ref{MatReg3}) 
as the restriction of the algebra $C^{\infty}(M)$ onto the space of the zero modes.
The map can be written as $T_N(C^{\infty}(M)) = \Pi C^{\infty}(M) \Pi$
with the projection operator $\Pi$ onto the Dirac zero modes\footnote{
It is notable that this mathematical framework naturally arises in 
the context of the Tachyon condensation on non-BPS D-branes
\cite{Asakawa:2001vm,Terashima:2005ic}.
See also \cite{Asakawa:2018gxf, Terashima:2018tyi}.}.
The $N \times N$ matrix $T_N(f)$ for $f \in C^{\infty}(M)$ 
is called the Toeplitz operator of $f$. 

The Berezin-Toeplitz quantization was further generalized 
in \cite{Hawkins:1997gj,Hawkins:1998nj}
and applied to $U(1)$ charged scalar fields on $M$ \cite{Adachi:2020asg},
towards understanding the fuzzy description of D-branes\footnote{
See \cite{Nair:2020xzn} for a generalization to matrix valued scalar fields
and \cite{Hasebe:2017myo,Ishiki:2018aja} for the quantization using 
instanton configurations.}.
When $M$ has a nontrivial magnetic flux, charged scalar fields 
cannot be globally defined. They are defined on each local coordinate patch 
and glued together by a gauge transformation on any overlap 
of two patches. Such fields (mathematically called 
local sections of a complex line bundle) are naturally mapped to 
rectangular $N\times N'$ matrices, where the difference $N-N'$ 
is kept fixed to be the charge of the fields.
For a charged field $\varphi$ with charge $Q$, let us write its Toeplitz operator 
as $T_{NN'}(\varphi)$, which is $N\times N'$ matrix with $N-N'=Q$.  
With an appropriate construction which we will review later, 
it was shown that the the operator satisfies  
\cite{Hawkins:1997gj,Hawkins:1998nj}
\begin{align}
\lim_{N\to \infty}|T_{N}(f)T_{NN'}(\varphi) -T_{NN'}(f \varphi)| =0,
\label{quantization for module structure}
\end{align}
for any $f \in C^\infty(M)$ and a similar equation also holds for 
the left action of $T_{N'}(f)$ onto the rectangular matrix
$T_{NN'}(\varphi)$.
This is a generalization of the equation (\ref{MatReg2}) and 
shows that the $C^\infty(M)$-module structure of charged fields
can be approximated by the $M_N(\mathbb{C})$- and 
$M_{N'}(\mathbb{C})$- module structures of the rectangular matrices.

In this paper, we further investigate the Berezin-Toeplitz quantization by 
extending the work \cite{Adachi:2020asg}. 
We consider a more general setup than \cite{Adachi:2020asg}, such that
the scalar fields to be regularized take values in a general representation 
of an arbitrary gauge group. 
We will show that the regularization for such fields can also be achieved by 
rectangular matrices. We will then derive a general large-$N$ asymptotic expansion of the product of two Toeplitz operators up to the second order in $1/N$.
This expansion basically contains all important information of the quantization map and
the fundamental relations such as (\ref{MatReg1}), (\ref{MatReg2}) and (\ref{quantization for module structure}), can also be derived from this expansion.
By using the asymptotic expansion, 
we then construct an operator acting on the rectangular 
matrices such that its spectrum approaches in the commutative limit
to that of the continuum Laplacian on $M$ with an arbitrary configuration of 
the background gauge field.

This paper is organized as follows. 
In section 2, we first review the Berezin-Toeplitz quantization for scalar fields in a 
general gauge field background and then derive the asymptotic expansion.
In section 3, we construct the fuzzy Laplacian and show some examples of 
this construction.
In section 4, we summarize our results.


\section{Berezin-Toeplitz quantization}

In this section, we consider the Berezin-Toeplitz quantization of  scalar fields in the presence 
of nontrivial background gauge fields
\cite{Bordemann:1993zv, Ma-Marinescu,Hawkins:1997gj,Hawkins:1998nj,Hawkins:2005} (See also \cite{Adachi:2020asg}). After defining the quantization map, 
we derive the large-$N$ asymptotic expansion for Toeplitz operators.

\subsection{Berezin-Toeplitz quantization of scalar fields}

Let $M$ be a closed Riemann surface with a metric $g$. We denote by $\omega$ the 
volume form of $g$. Since $\omega$ is a nondegenerate closed 2-form, it is also a 
symplectic form on $M$. 

We denote by $L$ a complex line bundle with a particular $U(1)$ connection 
$A$ such that its field strength $F$ is proportional to the symplectic form as
\beq
    F = dA = \omega/V.
    \label{symplectic potential}
\eeq
Here, $V$ is the volume, $V=\frac{1}{2\pi}\int_M \omega$, so that 
$\frac{1}{2\pi}\int_M F =1$. 
The line bundle $L$ becomes very important below and will be used to realize 
the desired large-$N$ expansion satisfying (\ref{MatReg1})--(\ref{MatReg3}) or 
(\ref{quantization for module structure}).
The gauge field $A$ may be different from the physical background 
gauge field introduced below.\footnote{
The work \cite{Adachi:2020asg} treats the special case in which $A$ is identical to the 
physical gauge field.}

We next introduce physical gauge fields coupling to the scalar fields, 
to which we apply the Berezin-Toeplitz quantization.
We regard the scalar fields as sections of the vector bundle, ${\rm Hom}(E,E')$, 
and the gauge fields as its connection. Here, 
$E$ and $E'$ are arbitrary finite-rank vector bundles on $M$ with Hermitian 
inner products and Hermitian connections, and
${\rm Hom}(E,E')$ is the vector bundle on $M$ such that its fiber is given by a set 
of all linear maps from the fiber of $E$ to that of $E'$.\footnote{In this paper, 
we are mainly interested in the case where $E$ and $E'$ are 
bundles of representation spaces of a given gauge group. Another interesting 
case, which will be studied elsewhere, is such that $E$ and $E'$ are 
given as tensor products of $TM$ or $T^*M$. In this case, 
the sections of ${\rm Hom}(E,E')$ are not scalar but tensor fields. }
If the dimensions of the fibers of $E$ and $E'$ are $n$ and $n'$, respectively,
the fiber of ${\rm Hom}(E,E')$ is just a set of all $n' \times n$ matrices.
This definition of scalar fields covers all physically interesting cases. 
For example, when $E$ and $E'$ are 
given by $E=\tilde{L}^{\otimes n}$ and $E'=\tilde{L}^{\otimes m}$ 
with a certain complex line bundle $\tilde{L}$ with a $U(1)$ connection 
$\tilde{A}$, ${\rm Hom}(E,E')$ reduces to $\tilde{L}^{m-n}$. 
Sections of $\tilde{L}^{m-n}$ are just complex scalar fields coupled to the gauge field 
$\tilde{A}$ with the charge $m-n$. Another example is scalars fields in the 
adjoint representation of a non-abelian gauge group. 
By taking both $E$ and $E'$ to be the same as a vector bundle of 
the fundamental representation space of a given gauge group, 
sections of ${\rm Hom}(E,E')$ correspond to the adjoint scalars.
This definition of scalar fields in terms of ${\rm Hom}(E,E')$ 
is suitable for defining the quantization map, 
since there is a natural product of two scalar fields given by the 
composition of linear maps.
For two scalar fields $\varphi \in \Gamma({\rm Hom}(E,E'))$ and 
$\varphi' \in \Gamma({\rm Hom}(E',E''))$, where $\Gamma(E)$ denotes a set of
all sections of $E$, the pointwise composition of the linear maps on $M$ gives 
$\varphi' \varphi  \in \Gamma({\rm Hom}(E,E''))$.
This is the product that is to be promoted to the matrix product 
through the quantization map.

The quantization map is given in terms of the projection to Dirac zero modes
as briefly mentioned in the previous section. So let us introduce spinor fields on $M$. 
We consider the twisted spinor bundle, $S\otimes L^{\otimes N} \otimes E$,
where $S$ is the two-component spinor bundle on $M$, 
$N$ is a positive integer and $E$ is any Hermitian vector bundle.
We equip an inner product on $\Gamma(S\otimes L^{\otimes N} \otimes E)$ by
\beq
    \label{inner product}
    (\psi',\psi) := \int_M \omega \, (\psi')^{\dagger} \cdot \psi
\eeq
for $\psi,\psi' \in \Gamma(S\otimes L^{\otimes N} \otimes E)$.
Here, $\cdot$ is the inner product (contraction) of the all indices.
The norm on $\Gamma(S\otimes L^{\otimes N} \otimes E)$ is defined by 
$|\psi| = \sqrt{(\psi,\psi)}$.
We denote by $L^2(S\otimes L^{\otimes N} \otimes E)$ the subset of 
$\Gamma(S\otimes L^{\otimes N} \otimes E)$ given by all elements 
with finite norms.
Note that a scalar field $\varphi \in \Gamma ({\rm Hom}(E,E'))$ can be 
seen as a map from  $\psi \in \Gamma(S\otimes L^{\otimes N} \otimes E)$
to $\varphi \psi \in \Gamma(S\otimes L^{\otimes N} \otimes E')$, where the latter 
is defined as the pointwise product on $M$.
The quantization map is essentially given by the restriction of this action 
onto the Dirac zero modes, which we will discuss shortly.


We define the (twisted) Dirac operator $D^{(E)}$ as an elliptic differential 
operator on $\Gamma(S\otimes L^{\otimes N} \otimes E)$ given by
\beq
    \label{twisted Dirac}
    D^{(E)} \psi = i \gamma^{\alpha} \nabla_{\alpha}\psi,
\eeq
where $\{\gamma^{\alpha}\}$ are the gamma matrices in curved 
space satisfying $\{\gamma^{\alpha}, \gamma^{\beta} \} =2g^{\alpha\beta}$, namely,
for the constant gamma matrices $\{\gamma^{a}\}_{a=1,2}$
on a local orthogonal frame satisfying 
$\{\gamma^{a}, \gamma^{b} \} =2\delta^{ab}$, $\gamma^\alpha$ are given by 
$\gamma^\alpha = e^\alpha_a \gamma^a$ with $e^\alpha_a$ the inverse of 
the zweibein for the metric $g$.
The covariant derivative $\nabla_{\alpha}$ acts on 
$\psi \in \Gamma(S\otimes L^{\otimes N} \otimes E)$ as
\beq
    \label{covariant}
    \nabla_{\alpha} \psi = 
\left( \partial_{\alpha} + \Omega_{\alpha} - iNA_{\alpha} - i A^{(E)}_{\alpha}\right)\psi,
\eeq
where $\Omega_{\alpha}$ is the spin connection and $A^{(E)}_\alpha$ 
is the connection for the bundle $E$, which takes values in 
square matrices acting on the fiber of $E$.
We denote by $\mathrm{Ker}\,D^{(E)}$ the set of all normalizable zero modes 
of $D$ with respect to the inner product (\ref{inner product}).
As shown in appendix \ref{Vanishing}, $\mathrm{Ker}\,D^{(E)}$ becomes a 
$(d^{(E)} N + c^{(E)})$-dimensional vector space for sufficiently large $N$, where
$d^{(E)}$ and $c^{(E)}$ are the rank and the first Chern number of $E$, respectively.

By using the above structures, we can 
define the Berezin-Toeplitz quantization for scalar fields.
For any scalar field $\varphi \in \Gamma({\rm Hom}(E,E'))$, which gives a map 
$\Gamma(S\otimes L^{\otimes N} \otimes E) \rightarrow \Gamma(S\otimes L^{\otimes N} \otimes E')$,
the quantization map is defined by 
\beq
    T_{N}^{(E',E)} (\varphi) = \Pi' \varphi \Pi.
    \label{Toeplitz op}
\eeq
Here, $\Pi:\Gamma(S\otimes L^{\otimes N} \otimes E) \rightarrow 
\mathrm{Ker}\,D^{(E)}$ 
is the projection operator onto 
$\mathrm{Ker}D^{(E)}$ and $\Pi'$ is the similar projection for $E'$.
$T_{N}^{(E',E)} (\varphi)$ can be represented as a rectangular matrix with size 
$(d^{(E')} N + c^{(E')})\times (d^{(E)} N + c^{(E)})$ and is called the 
Toeplitz operator for $\varphi$. 
As we will see below,
the Toeplitz operator (\ref{Toeplitz op}) enjoys a nice large-$N$ 
asymptotic behavior, from which one can derive (\ref{MatReg1}), (\ref{MatReg2}) 
and (\ref{quantization for module structure}).

From (\ref{Toeplitz op}), we notice that the quantization map preserves 
the Hermitian conjugation as
\begin{align}
T_{N}^{(E,E')} (\varphi^\dagger)=(T_N^{(E',E)}(\varphi))^\dagger,
\label{preserving dagger}
\end{align}
where $\varphi^\dagger \in \Gamma({\rm Hom}(E',E))$ is the Hermitian conjugate 
of $\varphi$ defined by the inner product (\ref{inner product})
 and the $\dagger$ on the right-hand side is the 
Hermitian conjugate for the rectangular matrices.

\subsection{Asymptotic expansion of Toeplitz operators}
For any scalar fields $\varphi \in \Gamma({\rm Hom}(E,E'))$ and 
$\varphi' \in \Gamma({\rm Hom}(E',E''))$, let us consider their Toeplitz operators, 
$T(\varphi)=\Pi' \varphi \Pi$ and $T(\varphi)=\Pi'' \varphi' \Pi'$. 
Here and hereafter, we will omit all subscripts of the Toeplitz operators
as it is obvious from their arguments, and
we will recover the subscripts only when it may cause confusion.
The product $T (\varphi')T (\varphi)$ is a 
$(d^{(E'')} N + c^{(E'')})\times (d^{(E)} N + c^{(E)})$ matrix and has 
the following asymptotic expansion in $\hbar_N = V/N$:
\beq
    T (\varphi')T (\varphi) = \sum_{i=0}^{\infty} \hbar_{N}^i T (C_i(\varphi',\varphi)),
    \label{asym exp}
\eeq 
where $C_i: \Gamma({\rm Hom}(E',E'')) \otimes \Gamma({\rm Hom}(E,E'))
\rightarrow \Gamma({\rm Hom}(E,E''))$ represent bilinear differential operators such that the order of the 
derivatives in $C_i$ is at most $i$ for each argument. 
We find that the first three $C_i$'s are explicitly given by 
\als{
C_0(\varphi',\varphi) &= \varphi'\varphi,\\
C_1(\varphi',\varphi) &= -\frac{1}{2} (g^{\alpha\beta} +i W^{\alpha\beta})(\nabla_{\alpha}\varphi')(\nabla_{\beta}\varphi),\\
C_{2}(\varphi',\varphi)&=\frac{1}{8}(g^{\alpha\beta} +i W^{\alpha\beta})(\nabla_{\alpha}\varphi')
(R+4F_{12}^{(E')}) (\nabla_{\beta}\varphi) \\
& \quad +\frac{1}{8}(g^{\alpha\beta} +i W^{\alpha\beta})(g^{\gamma\delta} +i W^{\gamma\delta})
(\nabla_{\alpha}\nabla_{\gamma}\varphi')(\nabla_{\beta}\nabla_{\delta}\varphi).
\label{asymptotic exp}
}
Here, $R$ is the Ricci scalar and $W^{\alpha\beta} := \epsilon^{ab}e^\alpha_a e^\beta_b$,
which is the Poisson tensor induced by the symplectic structure.
$F_{12}^{(E')} = e_1^\alpha e_2^\beta F_{\alpha \beta}^{(E')}=
e_1^\alpha e_2^\beta (\partial_\alpha A^{(E')}_\beta-\partial_\beta A^{(E')}_\alpha-i[A^{(E')}_\alpha, A^{(E')}_{\beta}])$ is the curvature of
$E'$ in the orthonormal frame. The covariant derivatives in (\ref{asymptotic exp}) 
act on the scalar fields as 
\begin{align}
\nabla_\alpha \varphi = \partial_\alpha \varphi -iA_\alpha^{(E')} \varphi +i \varphi A_\alpha^{(E)},
\;\;\;
\nabla_\alpha \varphi' = \partial_\alpha \varphi' -iA_\alpha^{(E'')} \varphi' +i \varphi' A_\alpha^{(E')}.
\label{action of covariant derivatives}
\end{align}

We leave the proof of (\ref{asym exp}) to appendix \ref{asymptotic expansion}
(see also appendix~\ref{Consistency check of the asymptotic expansion} for 
a consistency check of our calculation),
and discuss here some important corollaries of (\ref{asym exp}).
From the leading term in (\ref{asym exp}), we first notice that
\als{
\label{generalized matrix regularization}
&\lim_{N\to\infty}\left|T(\varphi')T(\varphi)-T(\varphi'\varphi)\right|=0.\\
}
When both $E'$ and $E''$ are the trivial line bundle and 
$E=L^{\otimes(-Q)}$, the relation (\ref{generalized matrix regularization})
reduces to  (\ref{quantization for module structure}), as 
$\varphi' \in C^{\infty}(M)$ and $\varphi \in \Gamma(L^{\otimes Q})$.
When $E$ is also taken to be the trivial line bundle, it further reduces 
to (\ref{MatReg1}).

Next, suppose that four fields 
$\varphi_1 \in \Gamma({\rm Hom}(E,E'))$, 
$\varphi_2 \in \Gamma({\rm Hom}(E',E''))$, 
$\varphi_3 \in \Gamma({\rm Hom}(E,\tilde{E}'))$ and
$\varphi_4 \in \Gamma({\rm Hom}(\tilde{E}',E''))$ satisfy 
$\varphi_2\varphi_1 = \varphi_4 \varphi_3$. Then, from (\ref{asym exp}) 
we find that
\als{
\label{generalized matrix regularization 2}
&\lim_{N\to\infty}\left|\hbar_N^{-1}(T(\varphi_2)T(\varphi_1)-T(\varphi_4)T(\varphi_3))
+\frac{1}{2}T((g^{\alpha \beta}+iW^{\alpha \beta})((\nabla_\alpha \varphi_2)(\nabla_\beta \varphi_1)
- (\nabla_\alpha \varphi_4)(\nabla_\beta \varphi_3)     ))
\right|=0.\\
}
We further consider a special case in which $E'=E''$, $\tilde{E}'=E$, 
$\varphi_1=\varphi_4 =: \varphi \in {\rm Hom}(E,E')$, 
$\varphi_2 = f {\bf 1}_{E'} \in {\rm Hom}(E',E')$ and 
$\varphi_3 = f {\bf 1}_{E} \in {\rm Hom}(E,E)$, where $f \in C^{\infty}(M)$ and 
${\bf 1}_{E'}$ and ${\bf 1}_{E}$ are the identity matrices acting on the fibers of $E'$ and $E$, respectively. 
Then, (\ref{generalized matrix regularization 2}) reduces to
\als{
\label{generalized matrix regularization 3}
&\lim_{N\to\infty}\left|\hbar_N^{-1} [T(f{\bf 1}), T(\varphi)]^{(E',E)}_N
+iT^{(E',E)}_N(\{f, \varphi \})
\right|=0.\\
}
Here, we defined the generalized commutator,
\beq
\label{generalized commutator}
[T(f{\bf 1}), T(\varphi)]^{(E',E)}_N := T^{(E',E')}_N(f {\bf 1}_{E'}) T^{(E',E)}_N (\varphi)
- T^{(E',E)}_N (\varphi)T^{(E,E)}_N(f {\bf 1}_{E}),
\eeq
and the generalized Poisson bracket,
\beq
\label{generalized Poisson bracket}
\{f,\varphi \}:= W^{\alpha \beta}(\partial_{\alpha}f)(\nabla_{\beta}\varphi).
\eeq
If we put both $E$ and $E'$ to be the trivial line bundle and 
consider $\varphi$ as an ordinary function, 
the equation (\ref{generalized matrix regularization 3})
reduces to the second equation in (\ref{MatReg2}).


The equations (\ref{generalized matrix regularization}), 
(\ref{generalized matrix regularization 2}) and
(\ref{generalized matrix regularization 3}) for general vector bundles
are our new result. In particular, 
(\ref{generalized matrix regularization 3}) shows a new 
correspondence between the generalized Poisson bracket 
(\ref{generalized Poisson bracket}) and
the generalized commutator (\ref{generalized commutator}).
This correspondence is very useful in constructing the matrix Laplacian 
in the next section.

Before closing this section, we discuss a correspondence between the trace 
of matrices and the integration on $M$. For $\varphi \in \Gamma({\rm Hom}(E,E))$,
the Toeplitz operator $T(\varphi)$ is a square matrix. Its trace, ${\rm Tr}T(\varphi)$,
is related to the integral of the trace part of $\varphi$ as 
\begin{align}
\lim_{N\to \infty} \hbar_N \mathrm{Tr} \, T(\varphi) = 
\frac{1}{2\pi}\int_M \omega {\rm Tr}_E \varphi,
\label{trace-integral}
\end{align}
where ${\rm Tr}_E$ stands for the trace over the fiber of $E$.
See appendix~\ref{trace} for a proof of (\ref{trace-integral}).
Note that, when $E$ is the trivail line bundle, the relation (\ref{trace-integral}) 
reduces to (\ref{MatReg3}).
The relation (\ref{trace-integral}) also implies a correspondence for
 the inner product of scalar fields, as follows. 
For $\varphi, \varphi' \in \Gamma({\rm Hom}(E,E'))$,
there is the natural inner product,
\beq
    \label{inner product for scalars}
    (\varphi,\varphi') := 
    \frac{1}{2\pi}\int_M \omega \, {\rm Tr}_E 
    \left(\varphi^{\dagger} \varphi' \right).
\eeq
On the other hand, the Toeplitz operators behave as
\begin{align}
T(\varphi^\dagger)T(\varphi') = 
\sum_{i=0}^{\infty} \hbar_{N}^i T(C_i(\varphi^\dagger ,\varphi'))
= T(\varphi^\dagger \varphi') + O(1/N).
\end{align}
By taking the matrix trace on both sides and using
(\ref{preserving dagger}) and (\ref{trace-integral}),
we find that
\begin{align}
\lim_{N\to \infty} \hbar_N \mathrm{Tr} ( 
T(\varphi)^\dagger T(\varphi')) = (\varphi, \varphi').
\label{corresp for inner prod}
\end{align}
Thus, the inner product of the scalar fields is related to the 
Frobenius inner produt of their Toeplits operators.

\section{Laplacian for rectangular matrices}
In this section, we construct the matrix Laplacian, which is related,
via the Berezin-Toeplitz quantization, to the continuum Laplacian with 
a general background gauge field.
We will first show that the continuum Laplacian for a 
K\"{a}hler metric can be written in terms of isometric 
embedding functions and the generalized Poisson bracket 
(\ref{generalized Poisson bracket}).
Then, by using the relation 
(\ref{generalized matrix regularization 3}), 
we will find the corresponding operator on the matrix side.
We will also consider two examples, the fuzzy sphere and the fuzzy torus,
and show explicit forms of the matrix Laplacians.


\subsection{Laplacian and isometric embedding}

The Nash embedding theorem states that any Riemannian manifold can be isometrically embedded in the Euclidean space $\mathbb{R}^d$ for sufficiently large $d$.
Thus, for a closed Riemann surface $M$ with a metric $g$, 
there exists an isometric embedding 
\beq
X:M \to \mathbb{R}^d
\eeq
for sufficiently large $d$.
We denote the embedding coordinate functions as $\{X^A\}_{A=1,2,\cdots,d}$.
The word {\it isometric} means that the induced metric of the embedding 
is equal to the intrinsic metric $g$ on $M$:
\beq
(\partial_{\alpha} X^A)(\partial_{\beta} X^A) =g_{\alpha\beta},
\eeq
where the repeated index $A=1,2,\cdots,d$ is summed over.

Now, let us consider the Laplacian for the metric $g$.
For a scalar field $\varphi \in \Gamma({\rm Hom}(E,E'))$, 
the Laplacian is defined by
\beq
    \Delta \varphi := -g^{\alpha\beta} \nabla_{\alpha} \nabla_{\beta}
    \varphi,
    \label{def of Laplacian}
\eeq
where the covariant derivatives acts on $\varphi$ as 
(\ref{action of covariant derivatives}).
This Laplacian is a positive semi-definite Hermite operator with respect to the inner product (\ref{inner product for scalars}).
Below, we will prove that this operator can also be written by using 
the isometric embedding as
\beq
\label{laplacian embedding}
    \Delta \varphi = - \{X^A,\{X^A,\varphi \}\},
\eeq
where $\{\;\; , \;\; \}$ is the generalized Poisson bracket 
 (\ref{generalized Poisson bracket}).
We start from the right hand side of (\ref{laplacian embedding}) and calculate it
as follows:
\als{- \{X^A,\{X^A,\varphi \}\}&= -W^{\alpha\beta}W^{\gamma\delta}
(\partial_{\alpha}X^A)\nabla_{\beta}[(\partial_{\gamma}X^A)(\nabla_{\delta}\varphi)]\\
&=
-W^{\alpha\beta}W^{\gamma\delta}(\partial_{\alpha}X^A)[(\nabla_{\beta}\partial_{\gamma}X^A)(\nabla_{\delta}\varphi) +(\partial_{\gamma}X^A)(\nabla_{\beta}\nabla_{\delta}\varphi) ]\\
&=
-W^{\alpha\beta}W^{\gamma\delta}[\nabla_{\beta}\{(\partial_{\alpha}X^A)
(\partial_{\gamma}X^A)\}(\nabla_{\delta}\varphi)-(\nabla_{\beta}\partial_{\alpha}X^A)(\partial_{\gamma}X^A)(\nabla_{\delta}\varphi)\\
&\qquad +(\partial_{\alpha}X^A)(\partial_{\gamma}X^A)(\nabla_{\beta}
\nabla_{\delta}\varphi) ]\\
&=-W^{\alpha\beta}W^{\gamma\delta}[(\nabla_{\beta}g_{\alpha\gamma})
(\nabla_{\delta}\varphi)-(\nabla_{\beta}\partial_{\alpha}X^A)
(\partial_{\gamma}X^A)(\nabla_{\delta}\varphi)
+g_{\alpha\gamma}(\nabla_{\beta}\nabla_{\delta}\varphi) ]
\\
&=-W^{\alpha\beta}W^{\gamma\delta}g_{\alpha\gamma}(\nabla_{\beta}
\nabla_{\delta}\varphi)
\\
&=-g^{\beta\delta}(\nabla_{\beta}\nabla_{\delta}\varphi).
\label{siki henkei}
}
To obtain the first equality, we used the fact that $W^{\gamma\delta}$ is covariantly 
constant in two dimension.
In the fifth equality, we also used $\nabla_{\beta}g_{\alpha \gamma}=0$ 
and $W^{\alpha\beta}\nabla_{\beta}\partial_{\alpha}X^A = W^{\alpha\beta}
(\partial_{\beta}\partial_{\alpha}X^A-\Gamma^{\gamma}_{\alpha\beta}
\partial_{\gamma}X^A)=0$, where $\Gamma^{\gamma}_{\alpha\beta}$ is the 
Christoffel symbol. The last equality follows from the identity $W^{\alpha\beta}W^{\gamma\delta}g_{\alpha\gamma} = g^{\beta \delta}$, which follows from
 $W^{\alpha\beta} = \epsilon^{ab}e_a^\alpha e_b^\beta$.
The last expression in (\ref{siki henkei}) is just the Laplacian and thus, 
we have shown the equation (\ref{laplacian embedding}).

\subsection{Laplacians on fuzzy surfaces}
Now, let us consider the matrix counterpart of the Laplacian 
  (\ref{def of Laplacian}). 
For $\varphi \in \Gamma({\rm Hom}(E,E'))$,
the Toeplitz operator $T(\varphi)$ is a rectangular matrix 
with size $(d^{(E')}N + c^{(E')}) \times (d^{(E)}N + c^{(E)})$.
Let $B$ be any matrix of this size.   
From (\ref{generalized matrix regularization 3}) and 
(\ref{laplacian embedding}), we find that the continuum 
Laplacian is mapped to
\beq
\hat{\Delta}B := \hbar_N^{-2}[T(X^A{\bf 1}),[T(X^A{\bf 1}), B]].
\label{fuzzy matrix laplacian}
\eeq
Here, $[\;\; ,\;\; ]=[\;\; ,\;\; ]^{(E',E)}_N$ 
is the generalized commutator (\ref{generalized commutator}),
and we again omit the subscipts for simplicity.
Note that the operator (\ref{fuzzy matrix laplacian}) is
an positive semi-definite Hermite operator with respect to the Frobenius inner product. Below, we will argue that the spectra of the original and the regularized Laplacians 
agree with each other in the large-$N$ limit. 

Let $\{B_n\}$ be exact eigenstates of $\hat{\Delta}$ which
 satisfy
\begin{align}
\hat{\Delta}B_n = E_n B_n,  \;\;\;   \hbar_N{\rm Tr} (B_n^\dagger B_m) = \delta_{mn}.
\label{eigenstates of hatdelta}
\end{align}
The indices $m,n$ run from 1 to $(d^{(E')}N + c^{(E')})(d^{(E)}N + c^{(E)})$.
On the other hand, let $\{a_n\in \Gamma({\rm Hom}(E,E'))\}$ be exact eigenstates of $\Delta$ which satisfy 
\begin{align}
\Delta a_n = e_n a_n,  \;\;\;   (a_n, a_m) = \delta_{mn},
\label{eigenstates of deltaQ}
\end{align}
where the inner product is given by (\ref{inner product for scalars}).
Here, the indices run from 1 to infinity.
We focus on the eigenstates of $\hat{\Delta}$ which have eigenvalues of $O(N^0)$.
For such eigenstates, we write $E_n= \tilde{E_n} + \epsilon_n$, where 
$\tilde{E_n}= \lim_{N\rightarrow \infty}E_n$ and $\epsilon_n$ is the $1/N$
correction of $E_n$ satisfying $\lim_{N\rightarrow \infty}\epsilon_n =0$.
We will show that such eigenstates of $\hat{\Delta}$ are in one-to-one
correspondence with those of $\Delta$ in the large-$N$ limit.

First, we take a specific eigenstate $B_n$ with the eigenvalue 
$O(N^0)$ and write it as 
$B_n =T(b_n)$ by using a local section $b_n \in \Gamma({\rm Hom}(E,E'))$. 
This is always possible since the quantization map is surjective.
From (\ref{generalized matrix regularization 3}), we have 
\begin{align}
\hat{\Delta}B_n = T(\Delta b_n + \frac{1}{N}c_n),
\label{eigenmatrix}
\end{align}
where $c_n \in \Gamma({\rm Hom}(E,E'))$ is another section of $O(1)$ 
(The section $c_n$ is explicitly given as a combination consisting of 
$C_{i}(\cdot, \cdot)$, $X^A$ and $b_n$.).
Since the left-hand side of (\ref{eigenmatrix}) is equal to 
$E_n M_n$, we obtain 
\begin{align}
T(E_nb_n- \Delta b_n-\frac{1}{N}c_n) = 0.
\label{eq M1}
\end{align}
Here, notice that if $T(b_0)=0$ for a certain section $b_0$ of 
$O(1)$, $b_0$ goes to zero in the large-$N$ limit. This follows from the 
mapping between the trace and integral (\ref{trace-integral}). 
If $T(b_0)=0$, we have
\begin{align}
0&= \hbar_N{\rm Tr}\left( T(b_0)^\dagger T(b_0) \right)
\nonumber\\
&= \hbar_N{\rm Tr}T(b_0^\dagger b_0 + \frac{1}{N}C_1(b_0^\dagger, b_0) + \cdots )
\nonumber\\
&= \frac{1}{2\pi}\int_M \omega {\rm Tr}_E (b_0^\dagger b_0) +O(1/N).
\end{align}
In order for this equation to hold, $b_0$ has to vanish in the large-$N$ limit.
Thus,  (\ref{eq M1}) implies that 
\begin{align}
\lim_{N\rightarrow \infty } |E_n b_n- \Delta b_n-\frac{1}{N}c_n| = 0.
\label{eq M2}
\end{align}
Here, note also that $b_n$ is nontrivial and finite in the large-$N$ limit. 
This is because we have
\begin{align}
\frac{1}{2\pi}\int_M \omega {\rm Tr}_E (b_n^\dagger b_n) = 
\hbar_N{\rm Tr} (B_n^\dagger B_n) + O(1/N)
=1+O(1/N),
\end{align}
but this equation contradicts if $b_n=0$ or $\lim_{N\rightarrow \infty}|b_n|=\infty$.
Thus, $b_n$ should converge to a certain section $\tilde{b}_n$ in the large-$N$ 
limit. 
Furthermore, if we consider several different $n$'s, the sections 
$\tilde{b}_n$ satisfy the orthonormality condition. In fact, the large-$N$ limit
of the second equation in (\ref{eigenstates of hatdelta}) gives 
$(\tilde{b}_m, \tilde{b}_n ) = \delta_{mn}$.
The equation (\ref{eq M2}) then implies that
\begin{align}
 \Delta \tilde{b}_n =\tilde{E}_n\tilde{b}_n.
\end{align}
Thus, there exists an eigenstate of $\Delta$ with the 
eigenvalue $\tilde{E_n}= \lim_{N\rightarrow \infty}E_n$.
What we have shown above can be summarized as follows.
Let $I$ be any index set such that if $n \in I$, the eigenvalue 
$E_n$ is of $O(1)$. Then, for the set of orthonormal eigenstates 
$\{ (E_n, B_n) | n\in I  \}$ of $\hat{\Delta}$,
there always exists a corresponding set of orthonormal 
eigenstates $\{ (\tilde{E}_n, \tilde{b}_n) | n\in I  \}$ of $\Delta$.
The two set of eigenvlues are related by $\tilde{E_n}= \lim_{N\rightarrow \infty}E_n$.

We next focus on the converse of the above statement.
Namely, we start from the eigenstates $\{a_n\}$ of $\Delta$ 
and try to construct a corresponding eigenstates of $\hat{\Delta}$.
We define the Toeplitz operator of $a_n$ as
\begin{align}
B'_n := T(a_n). 
\end{align}
By applying $\hat{\Delta}$ on this equation and using 
(\ref{generalized matrix regularization 3}), we obtain
\begin{align}
\hat{\Delta}B'_n = T(\Delta a_n + \frac{1}{N}c'_n)
=e_n B'_n + \frac{1}{N} T(c'_n),
\label{eq M3}
\end{align}
where $c'_n$ is a section of $O(1)$.
This equation shows that in the large-$N$ limit, $B'_n$ becomes 
an eigenstate of $\hat{\Delta}$ with the eigenvalue $e_n$\footnote{A little more 
rigorous statement may be made as follows. We first expand $B'_n$ by using 
$B_n$ as $B'_n=\sum_{n'}q_{nn'}B_{n'}$. By substituting this into 
(\ref{eq M3}), multiplying $B_m^\dagger$ and taking the trace and 
the large-$N$ limit, we obtain $\lim_{N\rightarrow \infty}q_{nm}(e_n-E_m)=0$ 
for any $m$. If $e_n\neq \lim_{N\rightarrow \infty}E_m$ for all $m$, 
it leads to $q_{nm}\rightarrow 0$ for all $m$. This means $B'_n \rightarrow 0$, 
which contradicts with the orthonormality of $a_n$.
Thus, there exists at least one $E_m$ such that 
$\lim_{N\rightarrow \infty}E_m = e_n$.}.
The orthonrmality of $B'_n$ in the large-$N$ limit can also be shown 
in a similar way as we described above for $\tilde{b}_n$.
Thus, for any index set $I'$ and a set of orthonormal eigenstates 
$\{ (e_n, a_n) | n\in I'  \}$ of $\Delta$, 
we can construct a corresponding orthonrmal eigenstates 
$\{ (e_n, B'_n) | n\in I'  \}$ of $\hat{\Delta}$ in the large-$N$ limit.

The above arguments show that, in the large-$N$ limit, 
the $O(1)$ eigenvalues of $\hat{\Delta}$ are 
in one-to-one correspondence with those of $\Delta$.


\subsection{Laplacian on fuzzy $S^2$}
In this section, we consider the regularized Laplacian on fuzzy $S^2$ in 
a monopole background \cite{Madore:1991bw}. 
We consider the case in which 
$E=L^{\otimes(-Q)}$ and $E'$ is the trivial line bundle.
In this case, $\Gamma({\rm Hom}(E,E')) = \Gamma(L^{\otimes Q})$ and 
$(c^{(E)},d^{(E)}, c^{(E')}, d^{(E')})=(-Q, 1, 0,1)$.
The Toeplitz operator $T(\varphi)$ for $\varphi \in \Gamma(L^{\otimes Q})$
is thus a rectangular matrix of size $N\times (N-Q)$.

Let us consider $S^2$ in the standard polar coordinate $(\theta,\phi)\in [0,\pi]\times [0,2\pi)$.
We will focus on the chart $\mathcal{C}$ that does not include the north pole $\theta=0$ and the south pole $\theta = \pi$.
On $\mathcal{C}$, the standard metric and the symplectic form are defined by
\als{
    g &:= d\theta \otimes d\theta+ \sin^2 \theta d\phi\otimes d\phi,\\
    \omega &:= \sin \theta \, d\theta \wedge d\phi.
}
In this convention, the symplectic volume is $V=2$.
The connection of the line bundle $L$ satisfying 
(\ref{symplectic potential}) is given by
\beq
    A=\frac{1-\cos \theta}{2}d\phi.
\eeq
This is nothing but the Wu-Yang monopole configuration.
The standard isometric embeding of $S^2$ into $\mathbb{R}^3$ is given by 
\begin{align}
X^1=\sin\theta \cos\phi,\;\;\;
X^2=\sin\theta \sin\phi,\;\;\;
X^3=\cos\theta.
\end{align}

Now, let us consider a Laplacian acting on $\varphi \in \Gamma(L^{\otimes Q})$.
As mentioned above, this is the case where $E=L^{\otimes(-Q)}$ and $E'$ is the trivial line bundle. This means that $A^{(E)}=-QA$ and $A^{(E')}=0$.
Then, the Laplacian can be explicitly be written as
\als{
    \Delta \varphi 
    &= -\frac{1}{\sin\theta}\partial_{\theta}(\sin\theta\partial_{\theta}\varphi)
    -\frac{1}{\sin^2\theta}\partial_{\phi}^2\varphi
    +iQ\frac{1-\cos \theta}{\sin^2\theta}\partial_{\phi}\varphi 
    + \frac{Q^2}{2}\frac{1-\cos \theta}{\sin^2\theta}\varphi -\frac{Q^2}{4}\varphi.
}
The spectrum of this operator is exactly solvable using the monopole harmonics \cite{Wu:1976ge,Wu:1977qk}.
Let us define the following operators on $\mathcal{C}$:
\als{
	\label{angular momentum on local section}
	&\mathcal{L}^{(Q)}_1
	=i(\sin\phi\,\partial_\theta+\cot\theta\cos\phi\,\partial_\phi)
	-\frac{Q}{2}\frac{1-\cos\theta}{\sin\theta}\cos\phi,\\
	&\mathcal{L}^{(Q)}_2
	=i(-\cos\phi\,\partial_\theta+\cot\theta\sin\phi\,\partial_\phi)
	-\frac{Q}{2}\frac{1-\cos\theta}{\sin\theta}\sin\phi,\\
	&\mathcal{L}^{(Q)}_3
	=-i\partial_\phi- \frac{Q}{2}.
}
These operators corresponds to the angular momentum operators in the presence of a magnetic monopole with charge $Q/2$ located at the origin of a sphere.
They form a representation of the $\mathfrak{su}(2)$ algebra,
\beq
[\mathcal{L}^{(Q)}_A,\mathcal{L}^{(Q)}_B]=i\epsilon_{ABC}\mathcal{L}^{(Q)}_C,
\eeq
on the representation space $\Gamma(L^{\otimes Q})$.
A unitary irreducible representation of the $\mathfrak{su}(2)$ algebra is constructed by the highest weight method:
\als{
    &(\mathcal{L}^{(Q)}_A)^2 Y_{lm}^{(Q)} = l(l+1)Y_{lm}^{(Q)},\\
    &\mathcal{L}^{(Q)}_3 Y_{lm}^{(Q)} = mY_{lm}^{(Q)}.
}
Here, $\{Y_{lm}^{(Q)}| \, l = |Q|/2,|Q|/2+1,\cdots,\infty;m=-l,-l+1,\cdots,l \}$ are 
the monopole harmonics \cite{Wu:1976ge,Wu:1977qk}
and they form an orthonormal basis of the representation space 
$\Gamma(L^{\otimes Q})$.
By the direct calculation, we can show that the Laplacian is equal to the quadratic Casimir operator plus a constant:
\beq
\label{S^2 laplacian}
\Delta =(\mathcal{L}^{(Q)}_A)^2 - \frac{Q^2}{4}.
\eeq
Thus, the eigenvalues of $\Delta $ 
are $l(l+1)- \frac{Q^2}{4}$ and the eigenfunctions are given by $Y_{lm}^{(Q)}$.

Now, let us consider the regularized Laplacian
(\ref{fuzzy matrix laplacian}).
A direct calculation (for example in \cite{Adachi:2020asg,Ishiki:2019mvq}) shows that the embedding functions are mapped to
\begin{align}
    &T^{(E',E')}_N(X^A {\bf 1}_{E'}) = \frac{1}{J+1}L^{(J)}_A,  \quad
    T^{(E,E)}_N(X^A {\bf 1}_{E}) = \frac{1}{\tilde{J}+1}L^{(\tilde{J})}_A
\label{fuzzy s2 config}
\end{align}
where $J=(N-1)/2$, $\tilde{J}=(N-Q-1)/2$ and $L^{(J)}_A$ are the 
$(2J+1)$-dimensional representation 
of the $\mathfrak{su}(2)$ generators satisfying the Lie algebra,
\beq
[L^{(J)}_A,L^{(J)}_B]=i\epsilon_{ABC} L^{(J)}_C.
\eeq
The matrix configuration (\ref{fuzzy s2 config}) is 
known as the fuzzy sphere \cite{Madore:1991bw}.
For any $N \times (N-Q)$ matrix $B$, 
the regularized Laplacian (\ref{fuzzy matrix laplacian})
in this case is given by
\als{\label{fuzzy monopole 1}
\hat{\Delta}B &= \frac{N^2}{4}\left(\frac{1}{(J+1)^2} (L_A^{(J)})^2 B 
- \frac{2}{(J+1)(\tilde{J}+1)} L_A^{(J)} B L_A^{(\tilde{J})}+\frac{1}{(\tilde{J}+1)^2} B (L_A^{(\tilde{J})})^2\right)\\
&= \frac{N^2}{4}\left(\frac{J}{J+1}B+\frac{\tilde{J}}{\tilde{J}+1}B 
- \frac{2}{(J+1)(\tilde{J}+1)} L_A^{(J)} B L_A^{(\tilde{J})}\right),
}
where we used $(L_A^{(J)})^2 = J(J+1)$.

We then test whether the spectrum of $\hat{\Delta}$ agrees with that 
of the continuum Laplacian in the large-$N$ limit. Let us first introduce an operation,
\beq
    L_A \circ B := L_A^{(J)}B-B L_A^{(\tilde{J})}.
\eeq
Note that the operation $L_A\circ$ also forms $N(N-Q)$-dimensional representation of $\mathfrak{su}(2)$:
\beq
[L_A\circ,L_B\circ]=i\epsilon_{ABC}L_C \circ.
\eeq
It is known that there exist $N\times (N-Q)$ 
matrices called fuzzy spherical harmonics \cite{Grosse:1995jt,Baez:1998he,Dasgupta:2002hx, Dolan:2006tx, Ishiki:2006yr}, denoted by 
$\{\hat{Y}_{lm(J\tilde{J})}|\, l=|J-\tilde{J}|,|J-\tilde{J}|+1, \cdots,J+\tilde{J};\, m=-l,-l+1,\cdots,l\}$, which satisfy
\als{\label{fuzzy monopole harmonics}
&(L_A\circ)^2 \hat{Y}_{lm(J\tilde{J})} = l(l+1)\hat{Y}_{lm(J\tilde{J})},\\
&L_3 \circ \hat{Y}_{lm(J\tilde{J})} = m\hat{Y}_{lm(J\tilde{J})}.
}
These matrices are indeed the Toeplitz map of the 
monopole harmonics \cite{Adachi:2020asg}.
They are also a complete orthonormal basis of complex $N\times (N-Q)$ matrices.
The first equation of (\ref{fuzzy monopole harmonics}) implies that
\beq
\label{fuzzy monopole 2}
    L_A^{(J)} \hat{Y}_{lm(J\tilde{J})} L_A^{(\tilde{J})}
    =\frac{J(J+1)+\tilde{J}(\tilde{J}+1)-l(l+1)}{2}\hat{Y}_{lm(J\tilde{J})}.
\eeq
From (\ref{fuzzy monopole 1}) and (\ref{fuzzy monopole 2}), we find that
$\{\hat{Y}_{lm(J\tilde{J})}|\, l=|J-\tilde{J}|,|J-\tilde{J}|+1, \cdots,J+\tilde{J};\, m=-l,-l+1,\cdots,l\}$ are complete eigen modes of the operator $\hat{\Delta}$ and
the eigenvalues are given as
\als{
\hat{\Delta}\hat{Y}_{lm(J\tilde{J})} 
&= \frac{N^2}{4(J+1)(\tilde{J}+1)}\left(l(l+1)-\frac{Q^2}{4}\right) \hat{Y}_{lm(J\tilde{J})}\\
&=\left(l(l+1)-\frac{Q^2}{4} + O(N^{-1})\right) \hat{Y}_{lm(J\tilde{J})} .
}
Therefore, the spectrum indeed approaches the continuum spectrum 
 as $N$ goes to infinity.

\subsection{Laplacian on fuzzy $T^2$}
In this section, we consider the Laplacian on the fuzzy $T^2$ \cite{Connes:1997cr}.
We again consider the case in which 
$E=L^{\otimes(-Q)}$ and $E'$ is the trivial line bundle.

Let us consider a flat plane $\mathbb{R}^2$.
We define the metric and the symplectic form on $\mathbb{R}^2$ by
\als{g&:= dx^1 \otimes dx^1 + dx^2 \otimes dx^2,\\
\omega&:= dx^1 \wedge dx^2.
}
By introducing equivalence relations,
\beq
x^\alpha \sim x^\alpha + 2\pi \quad (\alpha =1,2),
\eeq
we define two-dimensional torus $T^2$ as the quotient space,
\beq
T^2 = \mathbb{R}^2/\sim.
\eeq
This space inherits the flat metric and the symplectic form on $\mathbb{R}^2$.
The symplectic volume of $T^2$ is then given by $V=2\pi$.
The $U(1)$ gauge field $A$ satisfying (\ref{symplectic potential})
is given by
\beq
A= \frac{1}{4\pi}(-x^2 dx^1 + x^1 dx^2),
\eeq
The embedding functions,
\begin{align}
X^1=\cos x^1, \;\;\;
X^2=\sin x^1,\;\;\;
X^3=\cos x^2,\;\;\;
X^4=\sin x^2,
\end{align}
gives an isometric embeding of $T^2$ into $\mathbb{R}^4$.

We then consider a Laplacian acting on $\Gamma(L^{\otimes Q})$,
where the background gauge fields are again taken to be
$A^{(E)}=-QA$ and $A^{(E')}=0$.
By employing the complex coordinate $z=\frac{x^1+i x^2}{\sqrt{2}}$, 
the Laplacian can be written as
\beq
\Delta \varphi =  -(\nabla_{z}\nabla_{\bar{z}}+\nabla_{\bar{z}}\nabla_{z}) \varphi
\eeq
for $\varphi \in \Gamma(L^{\otimes Q})$.
The commutator of $\nabla_{z}$ and $\nabla_{\bar{z}}$ produces the 
constant field strength multiplied by the charge $Q$. For $Q\neq 0$,
this commutation relation is identical to that of the 
creation and annihilation operators, up to some rescalings.
Indeed, if we introduce the creation and annihilation operators by
\beq
    \hat{a}:= i \sqrt{\frac{2\pi}{Q}} \nabla_{\bar z},\quad \hat{a}^{\dagger}:= i \sqrt{\frac{2\pi}{Q}} \nabla_{z},
\eeq
they satisfy the algebra $[\hat{a},\hat{a}^{\dagger}]=1$ on 
$\Gamma (L^{\otimes Q})$. 
In this case, we can write the Laplacian as
\beq
    \Delta \varphi  = \frac{Q}{\pi}\left(\hat{N}+\frac{1}{2}\right) \varphi,
\eeq
where $\hat{N}:=\hat{a}\hat{a}^{\dagger}$ is the number operator.
Therefore, the eigenvalues of $\Delta$ are the same as those of 
the 1-dimensional harmonic oscillator, $\frac{Q}{\pi}(n+\frac{1}{2})\,(n=0,1,\cdots)$. The eigenfunctions are explicitly computed in \cite{Adachi:2020asg} and they can be expressed in terms of the Jacobi-theta function and the Hermite polynomials.
On the other hand, for $Q=0$, 
the spectrum of the Laplacian is given by a sum of two integers 
which correspond to the momenta for the $x^{1}$ and $x^2$ directions.
Thus, the spectrum for $Q=0$ is completely different from those for $Q\neq 0$. 

Let us next consider the matrix Laplacian (\ref{fuzzy matrix laplacian}).
The explicit calculation in \cite{Adachi:2020asg} shows that the Toeplitz 
operators of the embedding functions are given by
\als{
    &T_N^{(E',E')}(X^1{\bf 1}_{E'}) = \frac{U^{(N)}+U^{(N)\dagger}}{2}, \;\;\;\;\;
    T_N^{(E',E')}(X^2{\bf 1}_{E'}) = \frac{U^{(N)}-U^{(N)\dagger}}{2i},\\
    &T_N^{(E',E')}(X^3{\bf 1}_{E'}) = \frac{V^{(N)}+V^{(N)\dagger}}{2},\;\;\;\;\;
    T_N^{(E',E')}(X^4{\bf 1}_{E'}) = \frac{V^{(N)}-V^{(N)\dagger}}{2i},
}
where
\als{U^{(N)} = e^{-\frac{\pi}{2N}}\left(
\begin{array}{ccccc}
     &  & & &1\\
     1&  & & &\\
     & 1 & &  &\\
     &  &\ddots & &\\
    &  & & 1& 
\end{array}
\right), \;\;\;\;
V^{(N)} =  e^{-\frac{\pi}{2N}}\left(
\begin{array}{cccc}
     q^{-1} &  & & \\
     & q^{-2} & & \\
     &  & \ddots&  \\
    &  & & q^{-N}
\end{array}
\right),
\label{clock-shift}
}
are the $N$-dimensional clock and shift matrices with $q=e^{i2\pi/N}$.
The Toeplitz operators $T^{(E,E)}_N(X^A{\bf 1}_{E})$ are given by 
replacing $N$ with $N-Q$ in the above expressions.
The matrices (\ref{clock-shift}) satisfy the will-known algebra
$U^{(N)}V^{(N)}=qV^{(N)}U^{(N)}$, which characterizes the 
fuzzy torus \cite{Connes:1997cr}.
The Laplacian (\ref{fuzzy matrix laplacian}) is then given by
\beq
\hat{\Delta}B = \frac{N^2}{4\pi^2} \left( U\circ U^{\dagger} \circ +V\circ V^{\dagger} \circ\right)B
\eeq
for any $N\times (N-Q)$ matrix $B$, where $A\circ B := A^{(N)}B-BA^{(N-Q)}$.
It is easy to see that for $Q=0$, 
the exact eigen modes of the Laplacian are given by $(U^{(N)})^m(V^{(N)})^n$,
where $m, n$ are integers. The corresponding eigenvalues 
approach to $m^2+n^2$ in the large-$N$ limit, which agree with 
the continuum spectrum.
On the other hand, for $Q\neq 0$, we could not obtain exact eigen modes 
for finite $N$. However, in \cite{Adachi:2020asg}, it is shown that the eigenvalue problem of the regularized Laplacian is equivalent to a class of Hofstadter problem \cite{1} and the problem was numerically solved. The result shows that 
the spectrum of the regularized Laplacian indeed agrees with the continuum Laplacian
in the large-$N$ limit.

.

\section{Summary}
In this paper, we proposed a general construction of  
Laplacians for scalar fields on fuzzy Riemann surfaces with a general 
background gauge field.
Our construction is based on the so-called Berezin-Toeplitz quantization,
which was first considered as a method of mapping 
commutative function algebra to noncommutative matrix algebra in such 
way that two algebraic structures of functions (the ordinary function algebra 
and the Poisson algebra) are well-approximated in terms of the matrix algebra.
We used a generalized form of the Berezin-Toeplitz quantization,
which can also be applied to fields in various representations of any gauge group.
The quantization map is given by (\ref{Toeplitz op}) and the 
fields are mapped to rectangular matrices in this quantization.
The Laplacian we constructed in this paper acts on those 
rectangular matrices and reproduces the continuum spectrum in the 
large-$N$ limit.

In order to construct the matrix Laplacian,
we first showed that the Toeplitz operators (\ref{Toeplitz op}) 
satisfy the asymptotic expansion (\ref{asym exp}).
In particular, this expansion implies the relation 
(\ref{generalized matrix regularization 3}), which 
 shows a mapping between the generalized Poisson bracket and the commutator-like operation for the Toeplitz operators. 

We then showed that any Laplacian for a K\"{a}hler metric on a Riemann surface
with an arbitrary background gauge field
can be written in terms of the isometric embedding function and the 
generalized Poisson bracket.
By using  (\ref{generalized matrix regularization 3}),
we mapped the continuum Laplacian on the Riemann surface to the matrix side. 
Thus, we obtained the general form of the matrix Laplacian (\ref{fuzzy matrix laplacian}).
We also argued that its spectrum indeed agrees with the original Laplacian 
in the large-$N$ limit.
We finally checked our construction for two examples of the fuzzy $S^2$ and 
the fuzzy $T^2$.

\section*{Acknowledgments}
The work of G. I. was supported, in part, 
by JSPS KAKENHI (Grant Number 19K03818).

\begin{appendix}
\numberwithin{equation}{section}
\setcounter{equation}{0}

\section{Vanishing theorem and index theorem}
\label{Vanishing}
In this appendix, for the Dirac operator $D^{(E)}$ on
$\Gamma(S\otimes L^{\otimes N} \otimes E)$, 
we will show that  $\mathrm{Ker}\, D^{(E)}$ is spanned by 
spinors with positive chirality and
${\rm dim}\mathrm{Ker}\, D^{(E)} = d^{(E)} N + c^{(E)}$ for sufficiently large $N$,
where $d^{(E)}$ and $c^{(E)}$ are the rank and the first Chern number of 
the vector bundle $E$.
The former statement is known as the vanishing theorem and the latter is a consequence of 
the index theorem. We also show that nonzero eigenvalues of 
$D^{(E)}$ has a large gap of $O(\sqrt{N})$. 
Below, we simply denote the Dirac operator by $D$, making the 
$E$-dependence implicit.

In two dimension, spinors can be decomposed according to their chirality: $\Gamma(S\otimes L^{\otimes N}\otimes E)=\Gamma^+(S\otimes L^{\otimes N}\otimes E)\oplus \Gamma^-(S\otimes L^{\otimes N}\otimes E)$.
If we take the gamma matrices in the orthonormal frame as the two 
Pauli matrices $\sigma^1$ and $\sigma^2$, then the chirality operator is 
given by $\sigma^3$. By adopting a basis where the chirality operator 
becomes diagonal, we can decompose $D$ as
\beq
    D=
    \left(
    \begin{array}{cc}
       0 & D^{-} \\
    D^{+} & 0
    \end{array}
    \right).
    \label{D D+ D-}
\eeq
Here, $\pm$ indicates the chirality of the space on which the operators are 
acting. This decomposition is always possible, since the Dirac operator anti-commute 
with the chirality operator.

We first show that $\mathrm{Ker} D^{-} = \{0\}$ for sufficiently large $N$, which 
means that $\mathrm{Ker}\, D^{(E)}$ is spanned by spinors with positive chirality.
We consider the square of $D$:
\beq
    D^2 = 
    \left(
    \begin{array}{cc}
        D^{-} D^{+} & 0 \\
        0 & D^{+}D^{-}
    \end{array}
    \right).
    \label{d square 1}
\eeq
We also use the Weitzenb$\ddot{\text{o}}$ck formula,
\beq
    \label{D^2}
    D^2 = -\nabla^{a} \nabla_{a} -(\hbar_N^{-1}+F^{(E)}_{12}) \sigma_3 + \frac{1}{4}R, 
\eeq
where $\nabla_a = e_a^\alpha \nabla_\alpha$, 
$\hbar_N = V/N$, $R$ is the scalar curvature and
$F^{(E)}_{12}=e_1^\alpha e_2^\beta F^{(E)}_{\alpha \beta}$ is the curvature 
of $E$ in the orthonormal frame.
By comparing (\ref{d square 1}) and (\ref{D^2}),
we find that
\beq
    D^{+} D^{-} = - \nabla^{a} \nabla_{a} + \hbar_N^{-1} +F^{(E)}_{12}+ \frac{1}{4}R.
\label{D+D-}
\eeq
By using this relation and also $(D^{+})^{\dagger} = D^{-}$, 
which follows from the Hermiticity of $D$, we obtain the following inequalities 
for all $\psi^- \in \Gamma^-(S\otimes L^{\otimes N} \otimes E)$:
\beq
    \label{eigenvalue inequality}
    | D^{-}  \psi^-|^2 = |\nabla_{a} \psi^-|^2 + \hbar_N^{-1} (\psi^-,\psi^-) + (\psi^-,(F^{(E)}_{12}+\frac{1}{4}R)\psi^-) \ge \left(\hbar_N^{-1} - C \right)|\psi^-|^2
    .
\eeq
Here, we introduced $C := |F^{(E)}_{12}+\frac{1}{4}R|$.
From the above inequalities, we conclude that 
$\mathrm{Ker} D^{-} = \{0\}$ for $\hbar_N^{-1} > C$
and this is indeed the case in the large-$N$ limit.

We next show that ${\rm dim}\mathrm{Ker}\, D = d^{(E)} N + c^{(E)}$
for sufficiently large $N$.
Note that, for sufficiently large-$N$, since $\mathrm{Ker} D^{-} = \{0\}$ as 
we saw above, we have the following relations:
\beq
\mathrm{dim} \, \mathrm{Ker} \, D = \mathrm{dim} \, \mathrm{Ker} \, D^{+} = \mathrm{Ind} \, D,
\eeq
where $\mathrm{Ind} \, D := \mathrm{dim} \, \mathrm{Ker} \, D^{+}
-\mathrm{dim} \, \mathrm{Ker} \, D^{-}$ is 
the analytical index of $D$.
By using the Atiyah-Singer index theorem, we obtain
\beq
\mathrm{dim} \, \mathrm{Ker} D= 
\mathrm{Ind} \, D = \frac{1}{2\pi}\int_M 
( NF {\rm Tr}_E({\bf 1}_E)  + {\rm Tr}_E F^{(E)}) = d^{(E)} N + c^{(E)},
\eeq
where ${\rm Tr}_E$ is the trace for the fiber of $E$ and 
${\bf 1}_E$ is the identity matrix on the fiber of $E$.
The coefficients are explicitly given by $d^{(E)}= {\rm Tr}_E({\bf 1}_E)$ and 
$c^{(E)}=\frac{1}{2\pi} \int_M {\rm Tr}_E F^{(E)}$.

Finally, we prove that nonzero eigenvalues of $D$ have a large gap 
of $O(\sqrt{N})$.
Let $\lambda$ be a non-zero eigenvalue of $D$ with the eigen spinor
$\psi \in \Gamma(S\otimes L^{\otimes N} \otimes E)$. 
We make the chirality decomposition as $\psi = \psi^+ \oplus \psi^-$, where 
$\psi^{\pm} \in \Gamma^{\pm}(S\otimes L^{\otimes N} \otimes E)$.
In terms of the expression (\ref{D D+ D-}), $\psi^+$ and $\psi^-$ are 
the upper and the lower components of $\psi$, respectively.
The eigenvalue equation for $D^2$ is then equivalent to
\beq
    \begin{cases}
        D^{-} D^{+} \psi^+ &= \lambda^2 \psi^+,\\
        D^{+} D^{-}  \psi^- &= \lambda^2 \psi^-.
    \end{cases}
\eeq
If $\psi^- \neq 0$, (\ref{eigenvalue inequality}) implies that 
$\lambda^2 \ge \hbar_N^{-1} -C$.
If $\psi^- = 0$, we have $\psi^+ \neq 0$ in order for $\psi$ to be nonzero.
By using the relation $D^{+} D^{-} (D^{+} \psi^+) = \lambda^2 (D^{+} \psi^+$), we again find that (\ref{eigenvalue inequality}) implies $\lambda^2 \ge \hbar_N^{-1} -C$. Thus, in any case, we have $\lambda^2 \ge \hbar_N^{-1} -C$.
This shows that $\lambda^2$ is of $O(N)$ and thus, the nonzero eigenvalues 
of $D$ indeed have a gap of $O(\sqrt{N})$.

\section{Assymptotic expansion for Toeplitz operators}
\label{asymptotic expansion}
In this appendix, we derive the large-$N$ asymptotic expansion (\ref{asym exp}).
The computation technique used in this appendix is based on \cite{Hawkins:2005}.

For $\varphi \in \Gamma({\rm Hom}(E,E'))$ and 
$\varphi' \in \Gamma({\rm Hom}(E',E''))$, let 
$T(\varphi)=\Pi' \varphi \Pi$ and $T(\varphi)=\Pi'' \varphi' \Pi'$
be their Toeplitz operators.
The product $T (\varphi')T (\varphi)$ can be written as
\als{
    \label{asymptotic1}
    T (\varphi') T(\varphi) &= \Pi'' \varphi' \Pi' \varphi \Pi \\
    &= T(\varphi'\varphi) - \Pi'' \varphi' (1-\Pi') \varphi \Pi.
}
We will compute the second term in the following.

In order to compute $1-\Pi'$, let us consider the following Hermite 
operator on $\Gamma(S\otimes L^{\otimes N} \otimes E')$:
\beq
    P^{(E')} := 
    \left(
    \begin{array}{cc}
        0 & D^{-} (D^{+} D^{-})^{-1} \\
        (D^{+} D^{-})^{-1} D^{+} & 0
    \end{array}
    \right),
\eeq
where $D^{\pm}$ are the off-diagonal elements of $D^{(E')}$ 
in the chiral decomposition (\ref{D D+ D-}). 
Note that, since $\mathrm{Ker}D^{-} = \mathrm{Ker}D^{+} D^{-} = \{0\}$ for sufficiently large $N$ as shown in appendix~\ref{Vanishing}, the inverse 
$(D^{+} D^{-})^{-1}$ always exists. Hereafter, we will omit the subscript $(E')$ 
and if we simply write $P$ or $D$, it shall mean $P^{(E')}$ or $D^{(E')}$,
respectively. The operator $P$ has the following properties:
\als{
\label{D-P identity}
&D P = P D,\;\;\;\;\; P D P = P.
}
The first identity implies that $\mathrm{Ker}(D P) = 
\mathrm{Ker}(P D) = \mathrm{Ker}D$. The second identity 
implies that $(D P)^2 = D P$, which together 
with the Hermiticity of $D P$, shows that  
$D P$ is a projection onto $(\mathrm{Ker}D)^{\perp}$, 
which is the orthogonal compliment of $\mathrm{Ker}D$.
This projection is nothing but $1-\Pi'$ and 
thus, we find the expression,
\beq
    \label{projection}
    1-\Pi' = D P = D P^2 D.
\eeq

We substitute (\ref{projection}) into (\ref{asymptotic1}),
and act it onto an arbitrary zero mode $\chi \in \mathrm{Ker}D^{(E)}$.
By taking the inner product with another zero mode 
$\psi \in \mathrm{Ker}D^{(E'')}$, we obtain
\als{
    \label{asymptotic2}
    (\psi,T(\varphi')T(\varphi) \chi) &= (\psi,T(\varphi' \varphi) \chi) - 
    (\psi, \varphi' D P^2 D \varphi \chi)\\
    &= (\psi,T(\varphi' \varphi) \chi) + (\psi, \dot{\varphi}' P^2 
    \dot{\varphi} \chi).
    }
Here, we introduced the notation $\dot{\varphi} := i\sigma^{a} (\nabla_{a} \varphi)$.
Because the Pauli matrices in $\dot{\varphi}$ flips the chirality, 
$\dot{\varphi} \chi$ has the negative chirality and accordingly 
$\dot{\varphi} \chi \in (\mathrm{Ker}D)^{\perp}$.
On $(\mathrm{Ker}D)^{\perp}$, the operator 
$1-\Pi' = D P$ acts as the identity operator.
This means that $P$ is the inverse of $D$ on 
$(\mathrm{Ker}D)^{\perp}$.
Consequently, (\ref{asymptotic2}) can be written as
\beq
    \label{asymptotic3}
    (\psi,T(\varphi')T(\varphi) \chi) = (\psi,T(\varphi'\varphi) \chi) 
    + (\psi,\dot{\varphi}' D^{-2} \dot{\varphi} \chi).
\eeq
We compute the operator $D^{-2}$ on $(\mathrm{Ker}D)^{\perp}$
as follows. First, from the Weitzenb$\ddot{\text{o}}$ck formula (\ref{D^2}),
we have
\begin{align}
D^2 = -2 \nabla_- \nabla_+ + (1-\sigma_3)\left( 
\hbar_N^{-1} + \frac{1}{2}R_1
\right),
\end{align}
where $\nabla_{\pm}:= \frac{1}{\sqrt{2}}(\nabla_{1} \pm i \nabla_2)$ and 
$R_1:= 2F_{12}^{(E')} +\frac{R}{2}$.
By taking the inverse of this on the negative chirality modes, we obtain
\begin{align}
 D^{-2}|_- &= (-2\nabla_- \nabla_+ +2\hbar_N^{-1} +R_1)^{-1}
\nonumber\\
 &= \frac{\hbar_N}{2} -\frac{\hbar_N}{2} (-2\nabla_- \nabla_+ +R_1)
 D^{-2}|_-.
\label{asymptotic4}
\end{align}
Here, we used the elementary identity, $(a+b)^{-1}=a^{-1}-a^{-1}b(a+b)^{-1}$.
The term $\nabla_- \nabla_+ D^{-2}|_-$ can be further evaluated 
by using the following commutation relation:
\als{
\label{asymptotic5}
[\nabla_+, D^2|_-] 
&= -2[\nabla_+, \nabla_-] \nabla_+ + (\nabla_+ R_1)\\
&= (2\hbar_N^{-1} + R_2) \nabla_+ + (\nabla_+ R_1)
}
where $R_2:= R- \frac{R}{2}\sigma_3 + 2 F^{(E')}_{12}$.
This commutation relation is equivalent to
\begin{align}
  &(D^2|_- + 2\hbar_{N}^{-1} + R_2) \nabla_{+} = \nabla_{+} D^2|_- - (\nabla_{+} R_1).
\end{align}
By multiplying $(D^2|_- + 2\hbar_{N}^{-1}  + R_2)^{-1}$ from the left
and $D^{-2}|_-$ from the right, we obtain
\als{
\nabla_{+} D^{-2}|_- &= (D^2|_- + 2\hbar_{N}^{-1} + R_2)^{-1}\nabla_{+} \\
& \quad - (D^2|_- + 2\hbar_{N}^{-1} + R_2)^{-1}(\nabla_{+} R_1)D^{-2}|_-.
}
Plugging this into (\ref{asymptotic4}), we obtain
\als{
\label{expansion of D}
    D^{-2}|_- &= \frac{\hbar_{N}}{2} - \frac{\hbar_{N}}{2} R_1 D^{-2}|_- + \hbar_{N} \nabla_- (D^2|_- + 2\hbar_{N}^{-1} + R_2)^{-1}\nabla_{+} \\
    & \quad - \hbar_{N} \nabla_- (D^2|_- + 2\hbar_{N}^{-1} + R_2)^{-1}(\nabla_{+} R_1)D^{-2}|_- .
}
By using $\nabla_{+} \psi =0 $ and $\nabla_{+} \chi=0$, 
we then obtain
\beq
    (\psi,T(\varphi')T(\varphi) \chi) = (\psi,T(\varphi'\varphi) \chi) 
    + \frac{\hbar_{N}}{2}(\psi, \dot{\varphi}' \dot{\varphi} \chi) + \epsilon,
    \label{TT}
\eeq    
where 
\als{
    &\epsilon := \epsilon_1 + \epsilon_2 + \epsilon_3,\\
    &\epsilon_1 :=- \frac{\hbar_{N}}{2} (\psi, \dot{\varphi}' R_1 D^{-2}|_- \dot{\varphi} \chi),\\
    &\epsilon_2 := -\hbar_{N} (\psi, (\nabla_- \dot{\varphi}') (D^2|_- 
    + 2\hbar_{N}^{-1} + R_2)^{-1} (\nabla_+ \dot{\varphi} )\chi), \\
    &\epsilon_3 := \hbar_{N} (\psi, (\nabla_- \dot{\varphi}') (D^2|_- 
    + 2\hbar_{N}^{-1} + R_2)^{-1}(\nabla_{+} R_1)D^{-2}|_- \dot{\varphi} \chi).
\label{epsilon123}
}
Let us estimate the order of $\epsilon$ with respect to $\hbar_{N}$. 
From general properties of the inner product and the norm, we find that
\als{
    &|\epsilon_1| \le \frac{\hbar_{N}}{2} |\psi|\, |\dot{\varphi}'|\, |R_1|\, |D^{-2}|_-|\,|\dot{\varphi}|\,|\chi|,\\
    &|\epsilon_2| \le \hbar_{N} |\psi| \, |\nabla_- \dot{\varphi}'| \, |(D^2|_- + 2\hbar_{N}^{-1} + R_2)^{-1}| \, |\nabla_+ \dot{\varphi}| \, |\chi|, \\
    &|\epsilon_3| \le \hbar_{N} |\psi| \, |\nabla_- \dot{\varphi}'| \, |(D^2|_- + 2\hbar_{N}^{-1} + R_2)^{-1}| \, |\nabla_+ R_1|\, |D^{-2}|_-| \, |\dot{\varphi}|\, |\chi|.
}
Note that $\dot{\varphi}', \dot{\varphi}, \nabla_- \dot{\varphi}', \nabla_+ \dot{\varphi}, R_1$ and $\nabla_+ R_1$ are all 
$N$-independent and hence their norms are finite in the large-$N$ limit.
In addition, we can normalize $\psi$ and $\chi$ in such a way that
their norms are $N$-independent.
The only objects with nontrivial $N$-dependence are
$D^{-2}|_-$ and $(D^2|_- + 2\hbar_{N}^{-1} + R_2)^{-1}$.
As we discussed in appendix~\ref{Vanishing}, all eigenvalues of $D^2|_-$ 
are in the range $[\hbar_{N}^{-1}- C, \infty)$, where $C$ is an 
$N$-independent constant. Hence, the eigenvalues of $D^{-2}|_-$ are 
in $(0,(\hbar_{N}^{-1} - C)^{-1}]$.
From this property and the fact that
the norm of a positive operator is equal to its maximum eigenvalues,
we find that
\beq
    |D^{-2}|_-| = {O}(\hbar_{N}).
\eeq
A similar analysis also leads to
\beq
    |(D^2|_- + 2\hbar_{N}^{-1} + R_2)^{-1}| = O(\hbar_{N}).
\eeq
From these estimations, it follows that
\beq
    |\epsilon_1| = O(\hbar_{N}^2),\ 
    |\epsilon_2| = O(\hbar_{N}^2), \ 
    |\epsilon_3| = O(\hbar_{N}^3).
\eeq    
Then, since $\epsilon \le |\epsilon|\le |\epsilon_1|+|\epsilon_2|+|\epsilon_3|$,
we conclude that $\epsilon$ is $O(\hbar_{N}^2)$ and we can write the equation (\ref{TT}) as
\beq
    (\psi,T(\varphi')T(\varphi) \chi) = (\psi,T(\varphi'\varphi) \chi) 
    + \frac{\hbar_{N}}{2}(\psi, \dot{\varphi}' \dot{\varphi} \chi) + O(\hbar_N^2).
\eeq
This is nothing but the first two terms of the 
asymptotic expansion (\ref{asym exp}).
By using the relation $\gamma^a \gamma^b = \delta^{ab} + i \epsilon^{ab}\sigma^3$, 
we find that $C_0(\varphi',\varphi)$ and $C_1(\varphi' ,\varphi)$ in this expansion
are indeed given by those in (\ref{asymptotic exp}).

We can further obtain $C_2(\varphi',\varphi)$ in the following manner.
The contribution of $O(\hbar_{N}^2)$ comes from $\epsilon_1$ and 
$\epsilon_2$.
As for $\epsilon_1$, the operator $D^{-2}|_-$ 
in  (\ref{epsilon123}) can be again expanded as in 
(\ref{expansion of D}) and only the first term of the 
right-hand side of (\ref{expansion of D}) 
contributes to $C_2(\varphi_1,\varphi_2)$.
Similarly, in estimating $\epsilon_2$, the operator $(D^2|_- + 2\hbar_{N}^{-1} + R_2)^{-1}$ is expanded as $\frac{\hbar_{N}}{4}+O(\hbar_N^2)$.
After a short calculation, one finds that
$C_2(\varphi_1,\varphi_2)$ is exactly given by the expression in (\ref{asymptotic exp}).
Note that by applying this calculation recursively, 
one can in principle obtain arbitrary higher order contributions 
of the asymptotic expansion.

\section{Consistency check of the asymptotic expansion}
\label{Consistency check of the asymptotic expansion}
In this appendix, we give a consistency check of the asymptotic expansion 
(\ref{asym exp}) with (\ref{asymptotic exp}), derived 
in appendix~\ref{asymptotic expansion}.

Our consistency check is about the associativity of the matrix product.
For $\varphi \in \Gamma({\rm Hom}(E,E'))$,  
$\varphi' \in \Gamma({\rm Hom}(E',E''))$ and
$\varphi'' \in \Gamma({\rm Hom}(E'',E'''))$, we must have
\beq
\left(T(\varphi'')T(\varphi')\right)T(\varphi) = T(\varphi'')\left(T(\varphi')T(\varphi)\right).
\label{matrix associativity}
\eeq
By substituting the expansion (\ref{asym exp}), 
the associativity imposes the condition,
\beq
\label{associativity conditions}
\sum_{i,j=0}^{\infty}\hbar_{N}^{i+j}T\left(C_{j}\left(C_{i}(\varphi'' ,\varphi'),\varphi \right)-C_{i}\left(\varphi'', C_{j}(\varphi',\varphi)\right)\right)=0.
\eeq
At each order of $\hbar_{N}$, the summand should be separately vanishing.
Furthermore, (\ref{corresp for inner prod}) implies that, if $T(\varphi)=0$
in the large-$N$ limit, we have $\varphi = 0$. 
Thus, the equation (\ref{associativity conditions})
provides an infinite tower of constraints for $C_i's$:
\begin{align}
\sum_{i=0}^n
C_{n-i}\left(C_{i}(\varphi'' ,\varphi'),\varphi \right)
-C_{i}\left(\varphi'', C_{n-i}(\varphi',\varphi) \right) = 0,
\label{constraints for C}
\end{align}
for $n=0,1,2, \cdots$.

We will check that 
our $C_0,C_1,C_2$ in (\ref{asymptotic exp}) indeed satisfy the conditions
 (\ref{constraints for C}) up to $n=2$, which corresponds 
 to the second order of $\hbar_{N}^2$ in (\ref{associativity conditions}). 
First, the left-hand side of (\ref{constraints for C}) for $n=0$ is 
given by
\als{
&C_{0}\left(C_0(\varphi'',\varphi'),\varphi \right)-
C_{0}\left(\varphi'',C_{0}(\varphi',\varphi)\right)
=(\varphi'' \varphi' )\varphi
-\varphi''(\varphi' \varphi).
}
This is vanishing because of the associativity of the linear maps on the 
fiber vector spaces.
Next, for $n=1$, the left-hand side of (\ref{constraints for C}) is given by
\als{
&\sum_{i=0}^1
C_{1-i}\left(C_{i}(\varphi'' ,\varphi'),\varphi \right)
-C_{i}\left(\varphi'', C_{1-i}(\varphi',\varphi) \right)
\\ 
&=-(\nabla_-(\varphi'' \varphi'))(\nabla_+\varphi)
+\varphi''(\nabla_-\varphi')(\nabla_+\varphi)
-(\nabla_-\varphi'')(\nabla_+ \varphi')\varphi
+(\nabla_-\varphi'')(\nabla_+(\varphi'\varphi)).
}
Here, we used the relation, $(g^{\alpha\beta} + i W^{\alpha\beta})(\nabla_{\alpha} A) (\nabla_{\beta} B) = 2(\nabla_{-} A)(\nabla_{+} B)$.
This is again vanishing because of the derivation property of the covariant 
derivatives.
Finally, for $n=2$, a long but straightforward calculation leads to
\als{
&\sum_{i=0}^2
C_{2-i}\left(C_{i}(\varphi'' ,\varphi'),\varphi \right)
-C_{i}\left(\varphi'', C_{2-i}(\varphi',\varphi) \right)
\\
&=(\nabla_- \varphi'')([\nabla_-,\nabla_+]\varphi')(\nabla_+\varphi) 
 - (\nabla_- \varphi'')(F_{12}^{(E'')}\varphi'-\varphi'F_{12}^{(E')})  (\nabla_+ \varphi).
}
This is also vanishing because 
$[\nabla_-,\nabla_+]\varphi'
=F_{12}^{(E'')}\varphi'-\varphi'F_{12}^{(E')}$.
Thus, our assymptotic expantion (\ref{asym exp}) with $C_i$'s given by
(\ref{asymptotic exp}) is consistent with the associativity 
condition (\ref{matrix associativity}) up to the second 
order of $\hbar_{N}^{2}$.

\section{Trace of Toeplitz operators}
\label{trace}
In this appendix, we prove the equation (\ref{trace-integral}).

Let $\{\psi_I | I=1,2, \cdots, d^{(E)}N+c^{(E)} \}$ 
be an orthonormal basis of $\mathrm{Ker}D^{(E)}$ satisfying 
$(\psi_I, \psi_J) = \delta_{IJ}$.
For $\varphi \in \Gamma({\rm Hom}(E,E))$, we write 
\begin{align}
\mathrm{Tr} \, T(\varphi) = \mathrm{Tr} (\Pi \varphi \Pi )
= \sum_I (\psi_I, \varphi \psi_I) =
\int_M \omega {\rm Tr}_{S\otimes E} (K^{(E)} \varphi). 
\label{trace with Bergmann Kernel}
\end{align}
Here, ${\rm Tr}_{S\otimes E}$ is the trace on the fiber of $S \otimes E$ and
$K^{(E)}$ is defined by
\begin{align}
K^{(E)}_{st}(x)= \sum_I (\psi_I(x))_s (\psi_I^{\dagger}(x))_t,
\end{align}
where $x\in M$ and $s,t$ are collective labels for the indices of $S \otimes E$.
$K^{(E)}$ corresponds to the diagonal elements of the so-called 
Bergmann Kernel of the Dirac operator $D^{(E)}$.
It is known that the Bergmann Kernel has the following 
large-$N$ asymptotic expansion \cite{asym},
\begin{align}
K^{(E)} = (2\pi \hbar_N)^{-1} P_+ {\bf 1}_E + O(N^0),
\end{align}
where ${\bf 1}_E$ is the identity matrix on the fiber of $E$ and 
$P_+:=(1+\sigma_3)/2$ is the projection onto the positive chirality modes of $S$.
By substituting this into (\ref{trace with Bergmann Kernel}), 
we can obtain (\ref{trace-integral}).

\end{appendix}


\end{document}